\newif\ifmulticol	\multicoltrue
\newif\ifshowgit	\showgittrue		% switches footer on/off
\newif\ifgitlocal	\gitlocalfalse		% use local file gitHeadLocal.gin
\newif\ifbiblatex	\biblatexfalse		% defaults to bibtex if false
\newif\ifbibnum		\bibnumfalse 		% num => superscripts, otherwise auth date
\newif\iflineno		\linenofalse
\newif\iftoc		\toctrue

\newif\iflucida		\lucidafalse
\newif\ifcm			\cmtrue
\newif\iflibertine	\libertinefalse
\newif\ifcharter	\charterfalse

\newcommand*{\secnumdepth}{0}			% sets secnumdepth in \mymaketitle
\newcommand*{\tocdepth}{1}				% sets tocdepth in \mymaketitle
\newcommand*{\reftitle}{References}		% title for references section
\newcommand*{\mytitleskip}{-0.2in}		% change skip after title

\multicoltrue\showgittrue\gitlocaltrue\biblatexfalse\bibnumfalse

\newcommand*{\mydocfontsize}{\ifcharter11pt\else\iflibertine11pt\else11pt\fi\fi}
\newcommand*{\setcol}{\ifmulticol twocolumn\else onecolumn\fi}

\documentclass[\mydocfontsize,\setcol]{article}

\newcommand*{\pdfauthor}{Steven A.~Frank}
\newcommand*{\pdftitle}{The Price equation program: simple invariances unify population dynamics, thermodynamics, probability, information and inference}
\input price.sty
\newcommand{\bdq}{\GD\mathbf{q}}
\newcommand{\bq}{\mathbf{q}}

\newcommand{\bda}{\GD\mathbf{a}}
\newcommand{\ba}{\mathbf{a}}
\newcommand{\bOne}{\mathbf{1}}

\newcommand{\bw}{\mathbf{w}}
\newcommand{\obq}{\mathbf{\dot{q}}}
\newcommand{\oq}{{\dot{q}}}

\newcommand{\bdz}{\GD\mathbf{z}}
\newcommand{\bz}{\mathbf{z}}

\newcommand{\bmm}{\mathbf{m}}

\newcommand{\zbar}{\bar{z}}

\newcommand{\abar}{\bar{a}}
\newcommand{\wbar}{\bar{w}}

\newcommand{\br}{\mathbf{r}}
\newcommand{\obr}{\mathbf{\dot{r}}}

\newcommand{\bF}{\mathbf{F}}
\newcommand{\bI}{\mathbf{I}}

\newcommand{\Gfb}{\boldsymbol{\Gf}}

\newcommand{\Gfbar}{{\bar{{\Gf}}}}

\newcommand{\LL}{\mathcal{L}}
\newcommand{\EE}{\mathcal E}
\newcommand{\KL}[2]{\mathcal D\left(#1||#2\right)}
\newcommand{\D}{\mathcal D}
\newcommand{\J}{\mathcal J}
\newcommand{\F}{\mathcal F}
\newcommand*{\Gx}{\xi}

\newcommand{\tr}{T}

\newcommand{\cov}{\mathrm{Cov}}
\newcommand{\fracdq}{\frac{\GD q_i}{q_i}}
\newcommand{\GDF}{\GD_{\raisebox{-2pt}{$\scriptstyle \mathrm{F}$}}}

\newcommand{\dlog}{\dd\log}
\newcommand{\bL}{\mathbf{L}}

\newcommand{\qt}{q_\Gth}

\newcommand{\topic}[1]{\bigskip\noindent\textit{\textbf{#1}}}

\newcommand*{\Ga}{\alpha}
\newcommand*{\Gb}{\beta}

\newcommand*{\GD}{\Delta}
\newcommand*{\Ge}{\epsilon}

\newcommand*{\Gk}{\kappa}
\newcommand*{\Gl}{\lambda}

\newcommand*{\Gm}{\mu}
\newcommand*{\Gom}{\omega}

\newcommand*{\Gs}{\sigma}

\newcommand*{\Gth}{\theta}
\newcommand*{\Gf}{\phi}

\DeclarePairedDelimiter\abs{\lvert}{\rvert}
\DeclarePairedDelimiter\norm{\lVert}{\rVert}
\DeclarePairedDelimiter\angb{\langle}{\rangle}
\DeclarePairedDelimiter\lrb{\lbrack}{\rbrack}
\DeclarePairedDelimiter\lr{\lparen}{\rparen}

\makeatletter
\let\oldabs\abs \def\abs{\@ifstar{\oldabs}{\oldabs*}}
\let\oldnorm\norm \def\norm{\@ifstar{\oldnorm}{\oldnorm*}}
\let\oldangb\angb \def\angb{\@ifstar{\oldangb}{\oldangb*}}
\let\oldlrb\lrb \def\lrb{\@ifstar{\oldlrb}{\oldlrb*}}
\let\oldlr\lr \def\lr{\@ifstar{\oldlr}{\oldlr*}}
\makeatother

\DeclareMathOperator{\E}{E}
\newcommand*{\dd}{\textrm{d}}

\newcommand*{\Eq}[1]{eqn~\ref{eq:#1}}

\newcommand*{\prt}{\partial}

\newcount\BoxNum \BoxNum 1\relax
\makeatletter
\newcommand*{\boxlabel}[1]{%
  \protected@write \@auxout {}{\string \newlabel {box:#1}{{\the\BoxNum}}{}}%
  \advance\BoxNum 1\relax}
\makeatother

\RequirePackage{booktabs} % For \toprule etc. in tables

\newcommand*{\mytitle}{\pdftitle}

\newcommand*{\myauthor}{Steven A.\ Frank}
\newcommand*{\myaddress}{Department of Ecology and Evolutionary Biology, University of California, Irvine, CA 92697--2525, USA}
\newcommand*{\mycontact}{\href{https://stevefrank.org}{https://stevefrank.org}}

\newcommand*{\mykeywords}{Natural selection, symmetry, maximum entropy, d'Alembert's principle, Bayesian inference}
\setabstract{-0.04}{\iftoc-1.3\else1.22\fi}{%
The fundamental equations of various disciplines often seem to share the same basic structure. Natural selection increases information in the same way that Bayesian updating increases information. Thermodynamics and the forms of common probability distributions express maximum increase in entropy, which appears mathematically as loss of information. Physical mechanics follows paths of change that maximize Fisher information. The information expressions typically have analogous interpretations as the Newtonian balance between force and acceleration, representing a partition between the direct causes of change and the opposing changes in the frame of reference. This web of vague analogies hints at a deeper common mathematical structure. I suggest that the Price equation expresses that underlying universal structure. The abstract Price equation describes dynamics as the change between two sets. One component of dynamics expresses the change in the frequency of things, holding constant the values associated with things. The other component of dynamics expresses the change in the values of things, holding constant the frequency of things. The separation of frequency from value generalizes Shannon's separation of the frequency of symbols from the meaning of symbols in information theory. The Price equation's generalized separation of frequency and value reveals a few simple invariances that define universal geometric aspects of change. For example, the conservation of total frequency, although a trivial invariance by itself, creates a powerful constraint on the geometry of change. That constraint plus a few others seem to explain the common structural forms of the equations in different disciplines. From that abstract perspective, interpretations such as selection, information, entropy, force, acceleration, and physical work arise from the same underlying geometry expressed by the Price equation. These claims of universal structure are, at present, conjectures that deserve further study.
}

\begin{document}

\mymaketitle

%% 1st parm is skip on left column at start of TOC, 2nd param is skip after TOC
\iftoc{\renewcommand{\baselinestretch}{0.82}\mytoc{-20pt}{\newpage}\renewcommand{\baselinestretch}{1}}\fi

\section{Introduction}

The Price equation is an abstract mathematical description for the change in populations. The most general form describes a way to map entities between two sets. That abstract set mapping partitions the forces that cause change between populations into two components, the direct and inertial forces. 

The direct forces change frequencies. The inertial forces change the values associated with population members. Changed values can be thought of as an altered frame of reference driven by the inertial forces. 

From the abstract perspective of the Price equation, one can see the same partition of direct and inertial forces in the fundamental equations of many different subjects. That abstract unity clarifies understanding of natural selection and its relations to such disparate topics as thermodynamics, information, the common forms of probability distributions, Bayesian inference, and physical mechanics.

In a special form of the Price equation, the changes caused by the direct and inertial forces cancel so that the total remains conserved. That conservation law defines a universal invariance and canonical separation of the direct and inertial forces. The canonical separation of forces clarifies the common mathematical structure of seemingly different topics. 

This article sketches the overall argument for the common mathematical structure of different subjects. The argument is, at present, a broad framing of conjectures. The conjectures raise many interesting problems that require further work. Consult \textcite{frank12naturalb,frank17universal} for mathematical details, open problems, and citations to additional literature.

\section{The abstract Price equation}

The Price equation describes the change in the average value of some property between two populations \autocite{price72extension,frank12naturalb}. Consider a population as a set of things. Each thing has a property indexed by $i$. Those things with a common property index comprise a fraction, $q_i$, of the population and have average value, $z_i$, for whatever we choose to measure by $z$. Write $\bq$ and $\bz$ as the vectors over all $i$. The population average value is $\zbar=\bq\cdot\bz=\sum q_iz_i$, summed over $i$. 

A second population has matching vectors $\bq'$ and $\bz'$. Those vectors for the second population are defined by the special set mapping of the abstract Price equation. In particular, $q_i'$ is the fraction of the second population derived from entities with index $i$ in the first population. The second population does not have its own indexing by $i$. Instead the second population's indices derive from the mapping of the second population's members to the members of the first population. 

Similarly, $z_i'$ is the average value in the second population of members derived from entities with index $i$ in the first population. Let $\GD$ be the difference between the derived population and the original population, $\GD\bq=\bq'-\bq$ and $\GD\bz=\bz'-\bz$. 

To calculate the change in average value, it is useful to begin by considering $q$ and $z$ as abstract variables associated with the first set, and $q'$ and $z'$ as corresponding variables from the second set. 

The change in the product of $q$ and $z$ is $\GD(qz) = q'z' - qz$. Note that $q'=q+\GD q$ and $z'=z+\GD z$. We can write the total change in the product as a discrete analog of the chain rule for differentiation of a product, yielding two partial change terms
\begin{align*}
  \GD(qz) &= (q+\GD q)(z+\GD z)-qz\\
          &= (\GD q)z+(q+\GD q)\GD z\\
          &= (\GD q)z+q'\GD z.
\end{align*}
The first term, $(\GD q)z$, is the partial difference of $q$ holding $z$ constant. The second term, $q'\GD z$, is the partial difference of $z$ holding $q$ constant. In the second term, we use $q'$ as the constant value because, with discrete differences, one of the partial change terms must be evaluated in the context of the second set.

The same product rule can be applied to vectors, yielding the abstract form of the Price equation 
\begin{equation}\label{eq:price}
  \GD\zbar=\GD(\bq\cdot\bz)=\bdq\cdot\bz+\bq'\cdot\bdz.
\end{equation}
The abstract Price equation simply partitions the total change in the average value into two partial change terms.

Note that $\bq$ has a clearly defined meaning as frequency, whereas $\bz$ may be chosen arbitrarily as any values assigned to members. The values, $\bz$, define the frame of reference. Because frequency is clearly defined, whereas values are arbitrary, the frequency changes, $\bdq$, take on the primary role in analyzing the structural aspects of change that unify different subjects. 

The primacy of frequency change naturally labels the first term, with $\bdq$, as the changes caused by the direct forces acting on populations. Because $\bq$ and $\bq'$ define a sequence of probability distributions, the primary aspect of change concerns the dynamics of probability distributions.

The arbitrary aspect of the values, $\bz$, naturally labels the second term, with $\bdz$, as the changes caused by the forces that alter the frame of reference, the inertial forces.

Table \ref{tbl:symbols} defines commonly used symbols. Tables \ref{tbl:equations1} and \ref{tbl:equations2} in Appendix B summarize mathematical forms and relations between disciplines.

\newcommand*{\trow}[3]{$#1$& \parbox[t]{10cm}{\raggedright\hangindent=15pt\hangafter=1 #2}& \ref{eq:#3}\\[0pt]}
\newcommand*{\ts}{\vskip4pt}
\newcommand*{\tbi}[1]{\textbf{\textit{#1}}}

\begin{table*}[t]
\caption{Definitions of key symbols and concepts}
\label{tbl:symbols}
\centering
%% \tablesize{} %% You can specify the fontsize here, e.g.,  \tablesize{\footnotesize}. If commented out \small will be used.
\begin{tabular}{clc}
\toprule
\textbf{Symbol}	& \textbf{Definition}	& \textbf{Equation}\\
\midrule

\trow{\bq}{Vector of frequencies with $\sum q_i=1$}{price}
\trow{\bz}{Values with average $\zbar=\bq\cdot\bz$; use $\bz\equiv\ba,\bF$, etc.\ for specific interpretations}{price}
\trow{\bdq}{Discrete changes, $\GD q_i=q_i'-q_i$, may be large}{price}
\trow{\obq}{Small, differential changes, $\bdq\rightarrow \obq\equiv\dd \bq$}{info}
\trow{\ba}{Relative change of the $i$th type, $a_i=\GD q_i/q_i\rightarrow\oq_i/q_i=\log q_i'/q_i$}{adefShort}
\trow{\bmm}{Malthusian parameter, $\bmm=\log \bq'/\bq$, log of relative fitness, $\bw$}{malthus}
\trow{\bw}{Relative fitness, $w_i=q_i'/q_i$, with $\bmm=\log \bw$}{adef}
\trow{\bF}{Direct nondimensional forces, may be used for values $\bz\equiv\bF$}{dalembert1}
\trow{\bI}{Inertial nondimensional forces, may be interpreted as acceleration (\ref{eq:Idef})}{dalembert1}
\trow{\Gfb}{Force vector $\bF\equiv\Gfb$ when specific for particular case}{action}
\trow{\bdq\cdot\bF}{Abstract notion of physical work as displacement multiplied by force}{info}
\trow{\KL{\bq'}{\bq}}{Kullback-Leibler divergence between $\bq'$ and $\bq$}{info}
\trow{\F}{Fisher information, nondimensional expression}{info}
\trow{\LL}{Lagrangian, used to find extremum subject to constraints}{action}
\trow{\bL}{Likelihoods, $L_\Gth$, for parameter values, $\Gth$; interpreted as force, $\bF\equiv\bL$}{likelihood1}
\trow{\GDF}{Partial change caused by direct forces, e.g., $\bdq\cdot\bF$ or $\bdq\cdot\Gfb$ or $\bdq\cdot\bL$}{Vw}
\trow{\norm{\cdot}}{Euclidean vector length, e.g., $\norm{\bz}$ or $\norm{\bF}$ or $\norm{\bdq}$}{normdef}
\trow{\br}{Unitary coordinates, $\br=\sqrt{\bq}$, with $\norm{\br}=1$ as invariant total probability}{unitary}

\bottomrule
\end{tabular}
\end{table*}

\section{Canonical form}

The prior section emphasized the primary role for the dynamics of probability distributions, $\bdq$, which follows as a consequence of the forces acting on populations. 

The canonical form of the Price equation focuses on the dynamics of probability distributions and the associated forces that cause change. To obtain the canonical form, define
\begin{equation}\label{eq:adefShort}
  a_i=\fracdq
\end{equation}
as the relative change in the frequency of the $i$th type.

We can use any value for $\bz$ in the Price equation. Choose $\bz\equiv\ba$. Then
\begin{equation}\label{eq:azero}
  \GD\abar=\bdq\cdot\ba+\bq'\cdot\bda=0,
\end{equation}
in which the equality to zero expresses the conservation of total probability
\begin{equation*}
  \abar=\bq\cdot\ba=\sum_iq_i\fracdq=\sum_i\GD q_i=0,
\end{equation*}
because the total changes in probability must cancel to keep the sum of the probabilities constant at one. 

Thus, \Eq{azero} appears as a seemingly trivial result, a notational spin on $\sum\GD q_i=0$. However, many generalities and connections between seemingly different disciplines follow from the partition of conserved probability into the two terms of \Eq{azero}. 

\section{Preliminary interpretation}

The Price equation by itself does not calculate the particular $\bdq$ values of dynamics. Instead, the equation emphasizes the fundamental constraint on dynamics that arises from invariant total probability. The changes, $\bdq$, must satisfy the constraint in \Eq{azero}, specifying certain properties that any possible dynamical path must have. 

Put another way, all possible dynamical paths will share certain invariant properties. It is those invariant properties that reveal the ultimate unity between different applications and disciplines. 

Note that $\bq$ is fundamental, whereas $\bz$ is an arbitrary assignment of value or meaning. The focus on $\bq$ corresponds to the reason why information theory considers only probabilities, without consideration of meaning or values. In general, the unifying fundamental aspect among disciplines concerns the dynamics of probability distributions. We can then add values or meaning to that underlying fundamental basis.

In particular, we can first study universal aspects of the canonical invariant form based on $\ba$. We can then derive broader results by simply making the coordinate transformation $\ba\mapsto\bz$, yielding the most general expression of the abstract Price equation in \Eq{price}. 

Constraints on $\zbar$ or $\GD\zbar$ specify additional invariances, which determine further structure of the possible dynamical paths and equilibria. Each $z_i$ may be a vector of values, allowing multiple constraints associated with the $\bz$ values. 

Alternatively, one can study the conditions required for $\GD\zbar$ to change in particular ways. For example, what are the necessary and sufficient patterns of association between initial frequency, $\bq$, relative frequency change, $\ba$, and value, $\bz$, to drive the change, $\GD\zbar$, in a particular direction?

\section{Temporal dynamics}

The frequency change terms, $\GD q_i$, arise from the abstract set mapping assignment of members in the second set to members in the first set. In some cases, the abstract set mapping may differ from the traditional notion of dynamics as a temporal sequence, in which $q_i'$ is the frequency of type $i$ in the second set. 

We may add various assumptions to achieve a temporal interpretation in which $i$ retains its meaning as a type through time. For example, following \textcite{price95the-nature}, we may partition $\bq\mapsto\bq'$ into two steps. In the initial step, $\bq\mapsto\bq^*$, the mapping preserves type, such that $q_i^*$ describes the frequency of type $i$ in the second set. 

In the subsequent step, $\bq^*\mapsto\bq'$, the mapping accounts for the forces that change type. For a force that makes the change $i\mapsto j$, we map type $j$ members in the second set to type $j$ members in the first set. Thus, $\GD q_j=q_j'-q_j^*$ describes the net frequency change from the gains and losses caused by the forces of type reassignment.

For this two-step process that preserves type, the net change $\bq\mapsto\bq'$ combines the type-changing forces with other forces that alter frequency. Thus, we may consider type-preserving maps as a special case of the general abstract set mapping. In this article, I focus on the properties of the general abstract set mapping.

\section{Key results}

Later sections use the abstract Price equation to show formal relations between natural selection and information theory, the dynamics of entropy and probability, basic aspects of physical dynamics, and other fundamental principles \autocite{frank17universal}. Here, I list some key results without derivation or discussion. This listing gives a sense of where the argument will go, providing a target for further development in later sections.

Throughout this article, I use ratios of vectors to denote elementwise division, for example $\bq'/\bq=q'_1/q_1,q_2'/q_2,\dots$. A constant added to or multiplied by a vector applies the operation to each element of the vector, for example, $a+b\bz$, for constants $a$ and $b$, yields $a+bz_i$ for each $i$. 

\topic{D'Alembert's principle of physical mechanics.} We can write the canonical Price equation of \Eq{azero} as d'Alembert's partition \autocite{frank15dalemberts,frank17universal} between the direct forces, $\bF=\ba$, and the inertial forces of acceleration, $\bI$, as
\begin{equation}\label{eq:dalembert1}
  \GD\abar=\lr{\bF+\bI}\cdot\bdq=0.
\end{equation}
This equation generalizes Newton's second law that force equals mass times acceleration, describing the balance between force and acceleration. Here, the direct forces, $\bF$, balance the inertial forces of acceleration, $\bI$, along the path of change, $\bdq$. The condition $\GD\abar=0$ describes conservative systems. For nonconservative systems, we can use $\ba\mapsto\bz$, with $\GD\zbar$ not necessarily conserved. 

\topic{Information theory.} For small changes, $\bdq\rightarrow\obq$ and $\bF=\ba\rightarrow\log\lr{\bq'/\bq}$, the direct force term is
\begin{align}
  \bdq\cdot\bF=\bdq\cdot\ba&=\KL{\bq'}{\bq}+\KL{\bq}{\bq'}\nonumber\\[3pt]
                           &=\sum\frac{\oq_i^2}{q_i}=\F,\label{eq:info}
\end{align}
in which $\D$ is the Kullback-Leibler divergence, a fundamental measure of information, and $\F$ is a nondimensional expression of Fisher information \autocite{cover91elements}.

\topic{Extreme action.} The term for direct force, or action, $\obq\cdot\bF$, yields frequency change dynamics, $\obq$, determined by the extremum of the action, subject to constraint 
\begin{equation}\label{eq:action}
  \LL=\sum\oq_i\Gf_i-\frac{1}{2\Gk}\lr{\sum\frac{\oq_i^2}{q_i}-C^2}-\Gx\lr{\sum\oq_i-0},
\end{equation}
in which $\Gfb=\bF$ is a given force vector. The first parenthetical term constrains the incremental distance between probability distributions to be $\F=\sum\oq_i^2/q_i=C^2$, for a given constant, $C$. The second parenthetical term constrains the total probability to remain invariant.

\topic{Entropy and thermodynamics.} The force vector, $\Gfb$, can be described as a growth process, $q'_i=q_ie^{\Gf_i}$, with $\Gf_i=\log\lr{q'_i/q_i}$. A constraint on the system's partial change in some quantity, $\obq\cdot\bz=B$, constrains the new frequency vector, $\bq'$. We may write the constraint as $\obq\cdot\log\bq'=-\Gl\lr{\obq\cdot\bz}=-\Gl B$, thus
\begin{equation*}
  \LL=-\obq\cdot\log\bq-\frac{1}{2\Gk}\lr{\F-C^2}-\Gx\lr{\obq\cdot\bOne-0}
  	-\Gl\lr{\obq\cdot\bz-B}.
\end{equation*}
The action term, $-\obq\cdot\log\bq$, is the increase in entropy, $-\bq\cdot\log\bq$. Maximizing the action maximizes the production of entropy. 

\topic{Maximum entropy and statistical mechanics.} In the prior example, the work done by the force of constraint is $\obq\cdot\bF_{\mathbf{c}}=-\Gl B$, with $\bF_{\mathbf{c}}=\log\bq'=\log k-\Gl\bz$. At maximum entropy, we obtain an equilibrium, $\log\bq'=\log\bq$. Thus, the maximum entropy equilibrium probability distribution is
\begin{equation}\label{eq:maxEnt}
  q=ke^{-\Gl z}.
\end{equation}
This Gibbs-Boltzmann-exponential distribution is the principal result of statistical mechanics. Here, we obtained that result through a Price equation abstraction that led to maximum entropy production, subject to a constraining invariance on a component of change in $\zbar$.

\topic{Constraint, invariance and sufficiency.} The maximum entropy probability distribution expresses the forces of constraint, $\bF_{\mathbf{c}}$, acting on $\bz$. Different constraints yield different distributions. For example, the constraint $\bq\cdot\lr{\bz-\Gm}^2=\Gs^2$ yields a Gaussian distribution for given mean, $\Gm$, and variance, $\Gs^2$. This constraint is sufficient to determine the form of the distribution. Similarly, for small changes, the total change of the direct forces
\begin{equation}\label{eq:fisherInfo}
  \bdq\cdot\ba=\bdq\cdot\bF\rightarrow\sum\frac{\oq_i^2}{q_i}=\F,
\end{equation}
does not require the exact form of the frequency changes, $\obq$. It is sufficient to know the Fisher information distance, $\sum\oq_i^2/q_i=\F$, which determines the subsets of the possible change vectors, $\obq$, with the same invariant Fisher distance, $\F$. Many results from the abstract Price equation express invariance and sufficiency.

\topic{Inference: data as a force.} Use $\Gth\equiv i$ as an index for different parameter values. Then $q_\Gth$ matches the Bayesian notion of a prior probability distribution for the values of $\Gth$. The posterior distribution is
\begin{equation}\label{eq:likelihood1}
  q_\Gth'=q_\Gth L_\Gth,
\end{equation}
in which the normalized likelihood, $L_\Gth$, describes the force of the data that drives the change in probability. In Price notation, the normalized likelihood is equivalent to the force vector, $\bL\equiv\bF$, and also $\bL-\bOne\equiv\ba$. With that definition for $\ba$ in terms of the force of the data, the structure and general properties of Bayesian inference follow as a special case of the abstract Price equation.  

\topic{Invariance, scale and probability distributions.} The maximum entropy probability distribution in \Eq{maxEnt} is invariant to affine transformation, $z\mapsto a+bz$, because $k$ and $\Gl$ adjust to $a$ and $b$. That affine invariance with respect to $z$, which arises directly from the abstract Price equation, is sufficient by itself to determine the structure of commonly observed probability distributions, without need of invoking entropy maximization. The structure of common probability distributions is
\begin{equation*}
  q=ke^{-\Gl e^{\Gb w}}.
\end{equation*}
The function $w(z)$ is a scale for $z$, such that a shift in that scale, $w\mapsto\Ga+w$, only changes $z$ by a constant multiple, and therefore does not change the probability pattern. Simple forms of $w$ lead to the various commonly observed continuous probability distributions. For example, $w(z)=\log z$ yields the stretched exponential distribution.

\section{History of earlier forms}

Before analyzing the abstract Price equation and the unification of disciplines, it is useful to write down some of the earlier expressions and applications of the Price equation from biology \autocite{frank95george,frank97the-price,frank12naturalb,walsh18evolution}. 

\subsection{Fitness and average excess}

This section extends the definition of relative changes in \Eq{adefShort}. Let $w_i=q_i'/q_i$ be the relative growth, or relative fitness, of the $i$th type. Then we may define
\begin{equation}\label{eq:adef}
  a_i=w_i-1=\frac{q_i'}{q_i}-1=\fracdq,
\end{equation}
which, in biology, is Fisher's average excess in fitness \autocite{fisher41average}. Note that $\GD q_i=q_ia_i$ and that the average value of $w$ is $\wbar=1$, thus $a_i=w_i-\wbar$. 

\subsection{Variance in fitness}

Considering $\ba$ as a measure of fitness, the first term of \Eq{azero} becomes the partial change in average fitness caused by the direct forces, $\bF$. In symbols
\begin{align}
  \GDF\mskip1mu\abar 
    =\bdq\cdot\ba&=\sum_i\GD q_i\lr{\frac{\GD q_i}{q_i}}\nonumber\\[3pt]
    &=\sum_i q_i\lr{\frac{\GD q_i}{q_i}}^2
  	=\sum_i q_ia_i^2=V_w,\label{eq:Vw}
\end{align}
in which $\GDF$ is the partial change caused by the direct forces, and $V_w$ is the variance in fitness.

\subsection{Fundamental theorem}

If we let 
\begin{equation*}
  a_i=\Ga x_i+\Ge_i
\end{equation*}
be the regression of fitness, $a_i$, on some predictor, $x_i$, and define $g_i=\Ga x_i$, then 
\begin{equation}\label{eq:ftns}
  \GDF\mskip1mu\abar 
    =\sum_i q_ia_i^2=V_g+V_\Ge.
\end{equation}
If one interprets $x_i$ as an inherited gene, and $\Ge_i$ as an environmental effect that is not transmitted to the next generation, then the partial change in fitness by natural selection that is transmitted to the next generation is $\GD_{\raisebox{-2pt}{$\scriptstyle \mathrm{NS}$}}\mskip1mu\abar=V_g$. This result is analogous to Fisher's fundamental theorem of natural selection \autocite{fisher58the-genetical,price72fishers,ewens89an-interpretation,frank97the-price}.

The analysis tracks three sets. The initial set before selection with $\abar$, the second set after selection with $\abar^\dagger$, and the third set after transmission with $\abar'$. The set after transmission retains only those changes associated with $x_i$, interpreted as an inherited gene, such that $\GD\abar=\abar'-\abar$. 

\subsection{Covariance form and replicators}

Using the definitions of relative fitness and average excess, the first term of the Price equation is
\begin{align}\label{eq:cov}
\begin{split}
  \bdq\cdot\bz&=\sum\lr{\GD q_i} z_i=\sum q_ia_iz_i\\
  &=\sum q_i(w_i-\wbar)z_i=\cov(w,z),
\end{split}
\end{align}
in which $\cov(w,z)$ is the covariance between fitness and value. This covariance implies that natural selection tends to increase the average value of $z$ in proportion to the association between fitness and value. If the values do not change, $\GD z_i=0$, then the total change is
\begin{equation*}
  \GD\zbar=\cov(w,z).
\end{equation*}
This covariance equation has been widely used to study natural selection \autocite{robertson66a-mathematical,wade85soft,gardner08the-price,queller17fundamental,walsh18evolution}. 

In one common application, sometimes referred to as the replicator problem, we label each individual in a population by its own unique index, $i$, and let $z_i=p_i$ be $0$ or $1$ to specify if each individual is a type $0$ or type $1$ individual \autocite{taylor78evolutionary,schuster83replicator}. We can think of $p_i$ as the frequency of type $1$ in individual $i$. Then $\bar{p}$ is the frequency of type $1$ individuals in the population, and 
\begin{equation}\label{eq:replicator}
  \GD\bar{p}=\cov(w,p)
\end{equation}
is the frequency change of types in the population \autocite{price70selection}. Here, we assume that individuals do not change their type during transmission, $\GD p_i=0$, so that the second Price equation term is zero. This assumption is usually interpreted in biology as the absence of mutation. 

\subsection{Levels of selection}

We can write the second Price equation term as
\begin{equation}\label{eq:secondterm}
  \bq'\cdot\bdz=\sum q_i'\lr{\GD z_i} = \sum q_iw_i\lr{\GD z_i} = \E\lr{w\GD z},
\end{equation}
in which $\E$ denotes the expectation operator for the average value. Combining this expression with \Eq{cov}, we obtain an alternative form of the Price equation
\begin{equation}\label{eq:priceStat}
  \GD\zbar=\cov(w,z)+\E\lr{w\GD z}.
\end{equation}
This form is often used to analyze how selection acts at different levels, such as individual versus group selection \autocite{price72extension,hamilton75innate}. As an example, consider a variant of the replicator problem, which uses $z\equiv p$, yielding
\begin{equation}\label{eq:groupSel}
  \GD\bar{p}=\cov(w,p)+\E\lr{w\GD p},
\end{equation}
in which $p_i$ now denotes the frequency of type $1$ individuals within the $i$th group of individuals, $w_i$ is the fitness of the $i$th group relative to all other groups, and $\GD p_i$ is the change in the frequency of type $1$ individuals within the $i$th group. Thus, the two terms can be interpreted as the change caused by selection between groups and the change caused by selection between individuals within groups.

\section{Mathematical properties}

This section illustrates mathematical properties of the Price equation. These mathematical properties set the foundation for unifying apparently different kinds of problems from different disciplines.

\subsection{Geometry and work}

Write the standard Euclidean geometry vector length as the square root of the sum of squares
\begin{equation}\label{eq:normdef}
  \norm{\bz}=\sqrt{\sum z_i^2}.
\end{equation}
For any vector $\bz$
\begin{equation*}
  \bdq\cdot\bz=\norm{\bdq}\norm{\bz}\cos\Gom=\cov(w,z),
\end{equation*}
in which $\Gom$ is the angle between the vectors $\bdq$ and $\bz$. If we interpret $\bz\equiv\bF$ as an abstract, nondimensional force, then
\begin{equation}\label{eq:work}
  \bdq\cdot\bF=\norm{\bdq}\norm{\bF}\cos\Gom
\end{equation}
expresses an abstract notion of work as the distance moved, $\norm{\bdq}$, multiplied by the component of force acting along the path, $\norm{\bF}\cos\Gom$. 

\subsection{Divergence between sets}

If we let $\bz\equiv\ba$ describe the relative growth of the various frequencies, $a_i=\GD q_i/q_i$, then the divergence between sets can be expressed as 
\begin{equation}\label{eq:diverge}
  \GDF\mskip1mu\abar=\bdq\cdot\ba
    =\sum\lr{\frac{\GD q_i}{\sqrt{q_i}}}^2=\norm{\frac{\bdq}{\sqrt{\bq}}}^2
    =V_w=R^2,
\end{equation}
in which $R$ is the radius of a sphere on which must lie all possible $\bdq\;/\mskip-1mu\sqrt{\bq}$ changes with the same divergence between sets. If we choose to interpret $\ba$ as an abstract notion of force, or fitness, acting on frequency changes, then $\bdq\cdot\ba$ is the work, with magnitude $\norm{\bdq\;/\mskip-1mu\sqrt{\bq}\,}^2$, that separates the probability distribution $\bq'$ from $\bq$. 

\subsection{Small changes, paths and logarithms}

If we think of the separation between sets as a sequence of small changes along a path, with each small change as $\bdq\rightarrow\obq$, then
\begin{equation*}
  \ba\rightarrow\frac{\obq}{\bq}=\dd\log\bq,
\end{equation*}
in which the overdot and the symbol ``$\dd$'' equivalently describe the differential. Then the partial change by direct forces separates the probability distributions of the two sets by the path length
\begin{equation}\label{eq:fisherDist}
  \GDF\mskip1mu\abar=\bdq\cdot\ba=\norm{\frac{\obq}{\sqrt{\bq}}}^2=\F,
\end{equation}
in which $\F$ is an abstract, nondimensional expression of the Fisher information distance metric.

\subsection{Unitary and canonical coordinates}

Let $\br=\sqrt{\bq}$. Then $\norm{\br}=1$, expressing the conservation of total probability as a vector of unit length, in which all possible probability combinations of $\br$ define the surface of a unit sphere. In Hamiltonian analyses of d'Alembert's principle for the canonical Price equation, $\br$ is a canonical coordinate system \autocite{frank15dalemberts}. 

\begin{figure*}[t]
\centering
\includegraphics[width=\hsize]{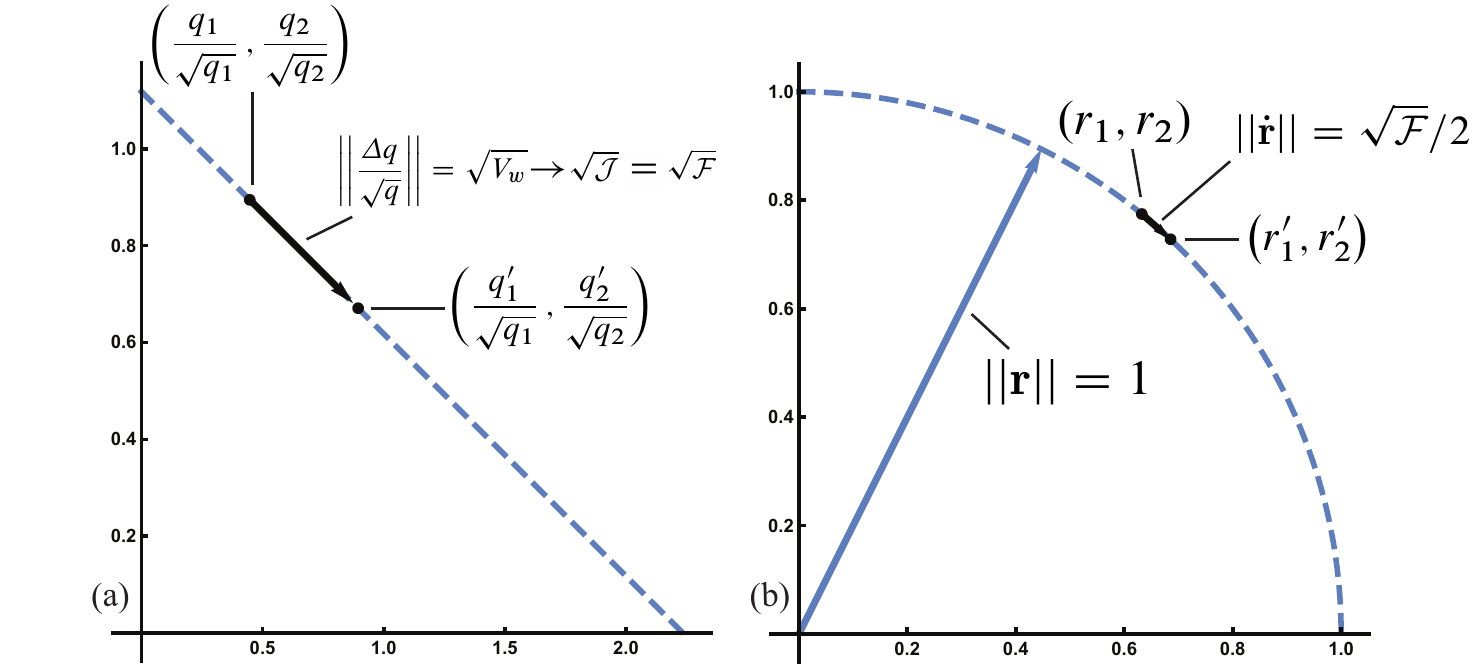}
\caption{Geometry of change by direct forces. See Table \ref{tbl:symbols} for definition of symbols. Tables \ref{tbl:equations1} and \ref{tbl:equations2} summarize distance expressions and point to locations in the text with further details. (\textbf{a}) The abstract physical work of the direct forces as the distance moved between the initial set with frequencies $\bq$, and the altered set with frequencies $\bq'$. For discrete changes, the frequencies are normalized by the square root of the frequencies in the initial set. The distance can equivalently be described by the various expressions shown, in which $V_w$ is the variance in fitness from population biology, $\J$ is the Jeffreys divergence from information theory, and $\F$ is the Fisher information metric which arises in many disciplines. The symbol ``$\rightarrow$'' denotes the limit for small changes. (\textbf{b}) When changes are small, the same geometry and distances can be described more elegantly in unitary square root coordinates, $\br=\sqrt{\bq}$.}
\label{fig:geometry}
\end{figure*}

The unitary coordinates, $\br$, also provide a direct description of Fisher information path length as a distance between two probability distributions
\begin{equation}\label{eq:unitary}
  4\norm{\obr}^2 = 4\norm{\dd\sqrt{\bq}}^2
  	=\norm{\frac{\obq}{\sqrt{\bq}}}^2=\F.
\end{equation}
The constraint on total probability makes square root coordinates the natural system in which to analyze Euclidean distances, which are the sums of squares. See Figure \ref{fig:geometry}. 

\subsection{Affine invariance}

Affine transformation shifts and stretches (multiplies) values, $\bz\mapsto a+b\bz$, for shift by $a$ and stretch by $b$. Here, addition or multiplication of a vector by a constant applies to each element of the vector. 

In the abstract Price equation
\begin{equation*}
  \GD\zbar=\bdq\cdot\bz+\bq'\GD\bz,
\end{equation*}
affine transformation, $\bz\mapsto a+b\bz$, alters the terms as: $\GD\zbar\mapsto b\GD\zbar$, because the shift constant cancels in the differences; $\bdq\cdot\bz\mapsto b\bdq\cdot\bz$, because in $\sum\lr{\GD q_i}\lr{a+bz_i}$, we have $\sum a\GD q_i=0$; and $\bq'\GD\bz\mapsto b\bq'\GD\bz$, because the shift constant cancels in the differences. The stretch factor $b$ multiplies each term and therefore cancels, leaving the Price equation invariant to affine transformation of the $\bz$ values. Much of the universal structure expressed by the Price equation follows from this affine invariance.

\subsection{Probability vs frequency}

In this article, I use \textit{probability} and \textit{frequency} interchangeably. Many subtle issues distinguish the concepts and applications associated with those alternative words. However, in this attempt to identify common mathematical structure between various subjects, those distinctions are not essential. See \textcite{jaynes03probability} for discussion.

\section{D'Alembert's principle}

The remaining sections repeat the list of topics in the \textit{Key results} section. Prior publications discussed these topics \autocite{frank12naturalb,frank17universal}. Here, I present additional details, roughly sketching how the structure provided by the abstract Price equation unifies various subjects.

We can rewrite the canonical Price equation for the conservation of total probability in \Eq{azero} as
\begin{equation}\label{eq:dalembert2}
  \GD\abar=\lr{\bF+\bI}\cdot\bdq=0.
\end{equation}
Here, $\bdq$ satisfies the constraint on total probability and any other specified constraints.  The direct forces are $\bF=\ba=\bdq/\bq$. The inertial forces are
\begin{equation}\label{eq:Idef}
  \bI=\frac{\GD^2\bq}{\bdq}-\frac{\bdq}{\bq},
\end{equation}
in which $\GD^2\bq=\GD(\bq'-\bq)$ is the second difference of $\bq$, which is roughly like an acceleration.

D'Alembert's principle is a generalization of Newton's second law, force equals mass times acceleration \autocite{lanczos86the-variational}. In one dimension, Newton's law is $F=-I$, for force, $F$, and mass times acceleration, $-I$, so that $F+I=0$. D'Alembert generalizes Newton's law to a statement about motion in multiple dimensions such that, in conservative systems, the total work for a displacement, $\bdq$, and total forces, $\bF+\bI$, is zero. Work is the distance moved multiplied by the force acting in the direction of the movement. 

The canonical Price equation of \Eq{azero} is an abstract, nondimensional generalization of d'Alembert for probability distributions that conserve total probability. The movement of the probability distribution between two populations, or sets, can be partitioned into the balancing work components of the direct forces, $\bdq\cdot\bF$, and the inertial forces, $\bdq\cdot\bI$. We can often specify the direct forces in a simple and clear way. The balancing inertial forces may then be analyzed by d'Alembert's principle \autocite{lanczos86the-variational}.

The movement of probability distributions in the canonical Price equation is always conservative, $\GD\abar=0$, so that d'Alembert's principle holds. When we transform to the general Price equation by $\ba\mapsto\bz$, then it may be that $\GD\zbar\ne0$ and the system is not conservative. In that case, we may consider constraints on $\GD\zbar$ and how those constraints influence the possible paths of change for $\bdq$. 

We can obtain a simple form of d'Alembert's principle for probability distributions when displacements are small, $\bdq\rightarrow\obq\equiv\dd\bq$. Define the relative change operator as $\dlog$, the differential of the logarithm. Then $\bF=\dlog\bq$ and $\bI=\dlog\lr{\dlog\bq}=\dlog^2\bq$, yielding
\begin{equation}\label{eq:dalembertLog}
  \lr{\bF+\bI}\cdot\dd\bq=\lr{\dlog\bq+\dlog^2\bq}\cdot\dd\bq=0,
\end{equation}
with the direct force proportional to the relative change in frequencies, and the inertial force proportional to the relative nondimensional acceleration in frequencies.

From \Eq{info}, the work of the direct forces, $\dd\bq\cdot\bF=\obq\cdot\bF=\F$, is the Fisher information path length that separates the probability distributions, $\bq'$ and $\bq$, associated with the two sets. The inertial forces cause a balancing loss, $\obq\cdot\bI=-\F$, which describes the loss in Fisher information that arises from the recalculation of the relative forces in the new frame of reference, $\bq'$. The balancing loss occurs because the average relative force, or fitness, is always zero in the current frame of reference, for example, $\bq\cdot\ba=\sum q_i (\oq_i/q_i)=0$. Any gain in relative fitness, $\obq\cdot\bF=\F$, must be balanced by an equivalent loss in relative fitness, $\obq\cdot\bI=-\F$.

Here, the notions of \textit{force,} \textit{inertia,} and \textit{work} are nondimensional mathematical abstractions that arise from the common underlying structure between the Price equation and the equations of physical mechanics. Similarly, the Fisher information measure here is an abstraction of the standard usage of the Fisher metric. 

By equating force with relative frequency change, we intentionally blur the distinction between external causes and internal effects. By describing change as the difference between two abstract sets rather than change through time or space, we intentionally blur the scale of change. By separating frequencies, $\bq$, from property values, $\bz$, we intentionally distinguish universal aspects of structural change between sets from the particular interpretations of property values in each application. The blurring of cause, effect and scale, and the separation of frequency from value, lead to abstract mathematical expressions that reveal the common underlying structure between seemingly different subjects. 

\section{Information theory}

When changes are small, the direct force term of the canonical Price equation expresses classic measures of information theory (\Eq{info}). In particular, $\obq\cdot\ba=\obq\cdot\bF$ is a symmetric expression of the Kullback-Leibler divergence, which measures the change in information associated with the separation between two probability distributions \autocite{cover91elements}. 

For small changes, the Kullback-Leibler divergence is equivalent to a nondimensional expression of the Fisher information metric. The Fisher metric provides the foundation for much of classic statistical theory and for the subject of information geometry \autocite{fisher25theory,amari00methods}. The Fisher metric also arises as an equivalent description for dynamics in many classic problems in physics and other subjects \autocite{frieden04science}.

What does it mean that the Price equation matches classic measures of information, which also arise other subjects? That remains an open question. I suggest that the Price equation reveals the common mathematical structure among those seemingly different subjects. That mathematical structure arises from the conserved quantities, invariances, or constraints that impose a common pattern on dynamics. By this interpretation, dynamics is just a description of the changes between a sequence of sets.

The key aspect of the Price equation seems to be the separation of frequencies from property values. That separation shadows Shannon's separation of the information in a message, expressed by frequencies of symbols in sets, from the meaning of a message, expressed by the properties associated with the message symbols. The Price equation takes that separation further by considering the abstract description of the separation between sets rather than the information in messages. \textcite{price95the-nature} was clearly influenced by the information theory separation between frequency and property in his discussion of a generalized notion of natural selection that might unify disparate subjects. 

The equivalence of the Price equation and information measures arises directly from the assumption of small changes. For larger changes, the relation between the Price equation and information remains an open problem. We might, for example, describe larger changes as
\begin{equation}\label{eq:malthus}
  q_i'=q_ie^{m_i},
\end{equation}
in which $m_i$ is a nondimensional expression for the total force that separates frequencies. From that expression,
\begin{equation}\label{eq:mdef}
  m_i=\log\frac{q_i'}{q_i}=\log w_i,
\end{equation}
in which $w_i$ is a form relative fitness, and $m_i$ is called the Malthusian parameter in biology. Then, similarly to \Eq{info}, we have
\begin{equation}\label{eq:jeffreys}
  \bdq\cdot\bmm=\KL{\bq'}{\bq}+\KL{\bq}{\bq'},
\end{equation}
which is known as the Jeffreys divergence. In this case, with $\bdq$ not necessarily small, we no longer have a direct equivalence to Fisher information. 

Information geometry, which analyzes continuous paths along contours of conserved total probability, describes the relations between Fisher information and this discrete divergence \autocite{dabak02relations}. The idea is that big changes, $\bdq$, become a series of small changes, $\obq$, along a continuous path that connects the endpoints, $\bq$ to $\bq'$. Each small step along the path can be described as a Fisher information path length, and the sum of those small lengths equals the Jeffreys divergence.

Earlier work in population genetics theory derived the total change caused by natural selection as $\sum\oq^2/q_i$ \autocite[reviewed by][]{ewens92an-optimizing,wei09pursuit,raju19a-variational}. That initial work did not emphasize the equivalence of the change by natural selection and Fisher information \autocite{frank09natural}. Here, the Fisher metric arises most simply as the continuous limiting form of the canonical Price equation description for the distance between two sets.

\section{Extreme action}

We can write \Eq{action} as
\begin{equation}\label{eq:actionV}
    \LL=\obq\cdot\Gfb-\frac{1}{2\Gk}\lr{\F-C^2}-\Gx\lr{\obq\cdot\bOne-0}.
\end{equation}
By the principle of extreme action, the dynamics, $\obq$, maximize or minimize (extremize) the action, $\obq\cdot\Gfb$, subject to the constraints. In this case, maximizing the action simply describes the fact that the movement, $\obq$, tends to be in the direction of the force vector, $\Gfb$, subject to any constraints on motion.

The Lagrangian, $\LL$, combines the action and the constraints into one expression. To illustrate the principle of extreme action with the Lagrangian above, we maximize the action subject to the constraints by solving $\prt\LL/\prt \oq_i=0$, while also solving for $\Gk$ and $\xi$ by requiring that $\F=C^2$ and $\obq\cdot\bOne=0$. The solution is
\begin{equation}\label{eq:actionDyn}
  \oq_i=\Gk q_i \lr{\Gf_i-\Gfbar},
\end{equation}
in which $\Gf_i-\Gfbar$ is the excess force relative to the average, and $\xi=\Gfbar$ follows from satisfying the constraint on total probability under the assumption of small changes. The constant, $\Gk=C/\Gs_\Gf$, satisfies the constraint on total path length, $\F=C^2$, in which $\Gs_\Gf$ is the standard deviation of the forces. We can rewrite the solution as
\begin{equation*}
  m_i=\frac{\oq_i}{q_i}=\Gk\lr{\Gf_i-\Gfbar}.
\end{equation*}
This expression shows that we can determine the frequency changes, $\obq$, from the given forces, $\Gfb$, or we can determine the forces from the given frequency changes. The mathematics is neutral about what is given and what is derived. 

In this case, $\Gfb$ is an arbitrary force vector. Using $\bz=\Gfb$ in the general Price equation does not necessarily yield $\GD\zbar=\GD\bar{\Gf}=0$. A nonconservative system does not satisfy d'Alembert's principle. Often, we can specify certain invariances associated with $\GD\zbar$, and use those invariances as additional forces of constraint on $\obq$ in the Lagrangian. The additional forces of constraint typically alter the dynamics and the potential equilibria, as shown in the following section.

Across many disciplines, problems can often be solved by this variational method of writing a Lagrangian and then extremizing the action subject to the constraints \autocite{lanczos86the-variational}. The difficulty is determining the correct Lagrangian for a particular problem. No general method specifies the correct form. 

In this example, the Price equation essentially gave us the form of the action and the constraints. Here, the action is the frequency displacement multiplied by the arbitrary force vector, $\obq\cdot\Gfb$, which is analogous to the physical work done in the movement of the probability distribution. The constraints follow from the conservation of total probability and the description of total distance moved as Fisher information, $\F$, which arises from the canonical Price equation. 

\section{Entropy and thermodynamics}

The tendency for systems to increase in entropy provides the foundation for much of thermodynamics \autocite{van-ness83understanding}. Entropy can be studied abstractly by the information entropy quantity, $\EE=-\bq\cdot\log\bq$. For small changes in frequencies, the change in entropy is $\dd\EE=-\obq\cdot\log\bq$. 

System dynamics often maximize the production of entropy \autocite{14beyond}. Maximum entropy production suggests that the dynamics may be analyzed by a Lagrangian in which the action to be maximized is the production of entropy, $-\obq\cdot\log\bq$.

In the basic Lagrangian for dynamics given by \Eq{actionV}, the action is the abstract notion of physical work, $\obq\cdot\Gfb$, the displacement, $\obq$, multiplied by the force, $\Gfb$.

The force vector, $\Gfb$, can be related to frequency change in a growth process, $q'_i=q_ie^{\Gf_i}$, with $\Gf_i=m_i=\log\lr{q'_i/q_i}$, as in \Eq{mdef}. The work becomes
\begin{equation}\label{eq:splitForce}
  \obq\cdot\Gfb=\obq\cdot\log\bq'-\obq\cdot\log\bq,
\end{equation}
in which the second term on the right is the production of entropy. 

If the system conserves the change in some quantity, $\GD\zbar=B$, then that invariant change imposes a constraint on the possible change in the probability distribution, $\obq=\bq'-\bq$. Suppose that the value $z_i$ is a property of a type, $i$, such that each type does not change its property value between sets, $\GD z_i=z_i'-z_i=0$. Then, from the general Price equation, $\GD\zbar=B$ implies $\obq\cdot\bz=B$. This constraint acts as a force that limits the possible probability distributions, $\bq'$, given the initial distribution, $\bq$.

We can express the constraint $\obq\cdot\bz=B$ on $\bz$ in terms of a constraint on $\bq'$ as $\log\bq'=\log k - \Gl\bz$, for constant, $k$. Then the constraint $\obq\cdot\bz$ has an equivalent expression in terms of $\bq'$ as
\begin{equation}\label{eq:dzconstraint}
  \obq\cdot\log\bq'= -\Gl\lr{\obq\cdot\bz}=-\Gl B.
\end{equation}
We can now split the total force, $\Gfb$, as in \Eq{splitForce} and, considering $\obq\cdot\log\bq'$ as a force of constraint, we can rewrite the Lagrangian of \Eq{actionV} as
\begin{equation}\label{eq:Lentropy}
  \LL=-\obq\cdot\log\bq-\frac{1}{2\Gk}\lr{\F-C^2}-\Gx\lr{\obq\cdot\bOne-0}
  	-\Gl\lr{\obq\cdot\bz-B}.
\end{equation}
The action term, $\dd\EE=-\obq\cdot\log\bq$, is the increase in entropy, $\EE=-\bq\cdot\log\bq$. Maximizing the action maximizes the production of entropy.

The maximization by solving $\prt\LL/\prt\oq_i=0$ subject to the constraints yields a solution with the same form as \Eq{actionDyn}. The force term is replaced by a partition of forces into components that match the direct entropy increase and the constraint on $\bz$ as
\begin{equation}\label{eq:forces}
  \Gf_i-\Gfbar=\EE_i^*-\Gl z_i^*,
\end{equation}
in which the star superscripts denote the deviations from average values, $\EE_i^*=-\log q_i-\EE$ and $z_i^*=z_i-\zbar$, thus
\begin{equation}\label{eq:entropyDyn}
  \oq_i=\Gk q_i\lr{\EE_i^*-\Gl z_i^*}.
\end{equation}
The value of $\Gk$ is $C/\Gs_\Gf$, as in the previous section. In this case, we use for $\Gf$ the partition of the forces on the right side of \Eq{forces} into the direct entropy and the constraining forces.

The constraint $\obq\cdot\bz=B$ implies
\begin{equation*}
  \Gl=\Gb_{\EE z}-\frac{B}{\Gk\Gs^2_z}.
\end{equation*}
The term $\Gb_{\EE z}$ is the regression of $-\log\bq$ on $\bz$, which acts to transform the scale for the forces of constraint imposed by $\bz$ to be on a common scale with the direct forces of entropy, $-\log\bq$. The term $B/\Gk\Gs^2_z$ describes the required force of constraint on frequency changes so that the new frequencies move $\zbar$ by the amount $\obq\cdot\bz=B$. The term $\Gs^2_z$ is the variance in $\bz$.

In these examples of dynamics derived from Lagrangians, the action is the partial change term of the direct forces derived from the universal properties of the Price equation. Thus, the maximum entropy production in this case can be interpreted as a universal partial maximum entropy production principle, in the Price equation sense of the partial change associated with the direct forces, holding the inertial frame constant \autocite{frank17universal}. 

In many applications, causal analysis reduces to this pattern of partial change by direct focal causes, holding other causes constant. The particular partition into direct, constraining, and inertial forces is a choice that we make to isolate or highlight particular causes \autocite{lanczos86the-variational}.

\section{Entropy and statistical mechanics}

When entropy reaches its maximum value subject to the forces of constraint, equilibrium occurs at $\bq'=\bq$. From the force of constraint given in the previous section, $\log\bq'=\log k-\Gl\bz$, the equilibrium can be written as
\begin{equation}\label{eq:expDistn}
  q=ke^{-\Gl z},
\end{equation}
in which I have dropped the $i$ subscript. This Gibbs-Boltzmann-exponential distribution is the principal result of statistical mechanics \autocite{feynman98statistical}. Here, we obtained the exponential distribution through a Price equation abstraction that led to maximum entropy production.

This result suggests that equilibrium probability distributions are simple expressions of maximum entropy subject to the forces of constraint. \textcite{jaynes57information,jaynes57informationII} developed this maximum entropy perspective in his quest to overthrow Boltzmann's canonical ensemble for statistical mechanics. The canonical ensemble describes macroscopic probability patterns by aggregation over a large number of equivalent microscopic particles.

The theory of statistical mechanics, based on the microcanonical ensemble, yields several commonly observed probability distributions. However, \textcite{jaynes03probability} emphasized that the same probability distributions commonly arise in economics, biology, and many other disciplines. In those nonphysical disciplines, there is no meaningful canonical ensemble of identical microscopic particles. According to Jaynes, there must another more general cause of the common probability patterns. The maximization of entropy is one possibility \autocite{frank09the-common}. 

Jaynes emphasized that increase in entropy is equivalent to loss of information. The inherent randomizing tendency in all systems causes loss of information. Maximum entropy is simply a consequence of that loss of information. Because systems lose all information except the forces of constraint, common probability distributions simply reflect those underlying forces of constraint. 

The Gibbs-Boltzmann-exponential distribution in \Eq{expDistn} expresses the simple force of constraint on the mean of some value, $\bz$, associated with the system. Different constraints lead to different distributions. For example, the constraint $\bq\cdot\lr{\bz-\Gm}^2=\Gs^2$ yields a Gaussian distribution for mean $\Gm$ and variance $\Gs^2$.

Jaynes invoked maximum entropy as a consequence of the thermodynamic principle that systems increase in entropy. Here, I developed the maximization of entropy from the abstract Price equation expression for frequency dynamics and the extreme action principle. 

Extreme action simply expresses the notion that changing frequencies align with the direction of the force vector. That geometric alignment is equivalent to the maximization of frequency change multiplied by force, an abstract notion of physical work. 

Jaynes argued that the fundamental notion of information sets the underlying structural unity of thermodynamics, probability, and many aspects of statistical inference. I argue for underlying unity based on abstract properties of invariance and geometry \autocite{frank17universal}. Those properties of invariance and geometry give a common mathematical structure to any problem that can be considered abstractly by the Price equation's description of the change between two sets. The next section reviews and extends these notions of invariance and common mathematical structure.

\section{Invariance and sufficiency}

The Price equation expresses constraints on the change in probability distributions between sets, $\bdq$. For example, if $\zbar$ is a constant, conserved value, then the changes, $\bdq$, must satisfy that constraint. We may say that the conserved value of $\zbar$ imposes a force of constraint on the frequency changes. This section relates the Price equation's abstract notions of change and constraint to Jaynes' arguments.

Jaynes emphasized that systems tend to increase in entropy or, equivalently, to lose information. Entropy increase is a force that drives a system to an equilibrium at which entropy is maximized subject to any forces of constraint. 

Because entropy increase is essentially universal, it is sufficient to know the particular forces of constraint to determine the most likely form of a probability distribution. Sufficiency expresses the forces of constraint in terms of conserved quantities.

Put another way, sufficiency partitions all possible populations into subsets. Each subset contains all of those populations with the same invariant conserved quantity. For example, if the constraint is a conserved value of $\zbar$, then all populations with the same invariant value of $\zbar$ fall into the same subset. 

To analyze the force arising from constraint on $\zbar$ and the most likely form of the associated probability distribution, it is sufficient to know that the dynamics of populations driven by entropy increase must remain within the subset with invariant values defined by the constraints of the conserved quantities.

Jaynesian thermodynamics follows from the general force of information loss, in which the constraints sufficiently describe the only information that remains after maximum information loss. 

The Price equation goes beyond Jaynes in revealing the underlying abstract mathematical structure that unifies seemingly different subjects. In all of the disciplines we have discussed, the key results for each discipline arise from the basic description of change between sets constrained by invariant conditions that we place on frequency, $\bq$, and value, $\bz$. In addition, the Price equation expresses the intrinsic invariance to affine transformation $\bz\mapsto a+b\bz$.

From the perspective of the abstract Price equation, notions of information and entropy increase arise as secondary descriptions of the underlying primary geometric aspects of change between sets subject to intrinsic invariances and to invariant conditions imposed as constraints. Those aspects of geometry and invariance set the shared foundations for many seemingly different disciplines.

\section{Inference: data as a force}

Jaynes considered information as a force that changes probability distributions. Entropy increase is the force that causes loss of information, driving probability distributions to maximum entropy subject to constraint. For inference, data provide an informational force that drives the Bayesian dynamics of probability distributions to provide estimates of parameter values. The parameters are typically the conserved, constrained quantities that are sufficient to define maximum entropy probability distributions.

How does the Jaynesian interpretation of data as an informational force in statistical inference follow from the underlying Price equation abstraction? Consider the estimation of a parameter, $\Gth$, such as the mean of an exponential probability distribution. In the Bayesian framework, we describe the current information that we have about $\Gth$ by the probability distribution, $\qt$. 

The value of $\qt$ represents the relative likelihood that the true value of the parameter is $\Gth$. The probability distribution over alternative values of $\Gth$ represents our current knowledge, or information, about $\Gth$. To relate this to the Price framework, note that we are now using $\Gth$ as the subscript for types instead of $i$. The vector $\bq$ now implicitly describes the set of values for $\qt$. 

Our problem concerns how new information about $\Gth$ changes the probability values to $\qt'$. The new probability values summarize the combination of our prior information in $\qt$ and the force of the new information in the data. This problem is the Bayesian dynamics of combining a prior distribution, $\qt$, with new data to generate a posterior distribution, $\qt'$, with $\GD\qt=\qt'-\qt$. 

We have from our universal definitions for change given earlier the relation $\qt'=\qt w_\Gth$, in which we called $w=q'/q$ the relative fitness, describing the force of change on probabilities. Here, the force arises from the way in which new data alters the net likelihood associated with a value of $\Gth$. 

Following Bayesian tradition, denote that force of the data as $\tilde{L}(D|\Gth)$, the likelihood of observing the data, $D$, given a value for the parameter, $\Gth$. To interpret a force as equivalent to relative fitness, the average value of the force must be one to satisfy the conservation of total probability. Thus, define
\begin{equation*}
  w_\Gth=L_\Gth=\frac{\tilde{L}(D|\Gth)}{\sum_\Gth \qt\tilde{L}(D|\Gth)}.
\end{equation*}
We can now write the classic expression for Bayesian updating of a prior, $\qt$, driven by the force of new data, $L_\Gth=L(D|\Gth)$, to yield the posterior, $\qt'$, as
\begin{equation}\label{eq:bayesUpdate}
  \qt'=\qt L_\Gth.
\end{equation}
By recognizing $\bL$ as a force vector acting on frequency change, we can use all of the general results derived from the Price equation. For example, the Malthusian parameter, $\bmm$, relates to the log-likelihood as
\begin{equation}\label{eq:logL}
  \bmm=\log\frac{\bq'}{\bq}=\GD\log\bq=\log\bL.
\end{equation}
This equivalence for log-likelihood relates frequency change to the Kullback-Leibler expressions for the change in information
\begin{equation}\label{eq:bayesJ}
  \GD\bq\cdot\log\bL=\KL{\bq'}{\bq}+\KL{\bq}{\bq'},
\end{equation}
which we may think of as the gain of information from the force of the data. Perhaps the most general expression of change describes the relative separation within the unitary square root coordinates as the Euclidean length
\begin{equation*}
  \GD\bq\cdot\bL=\norm{\frac{\GD\bq}{\sqrt{\bq}}}^2,
\end{equation*}
which is an abstract, nondimensional expression for the work done by the displacement of the frequencies, $\GD\bq$, in relation to the force of the data, $\bL$. 

I defined $\bL$ as a normalized form of the likelihood, $\skew{-4}\tilde{\bL}$, such that the average value is one, $\skew{-4}\bar{\bL}=\bq\cdot\bL=1$. Thus, we have a canonical form of the Price equation for normalized likelihood
\begin{equation}\label{eq:canonicalL}
  \GD\skew{-4}\bar{\bL}=\GD\bq\cdot\bL+\bq'\cdot\GD\bL=0.
\end{equation}
The second terms shows how the inertial forces alter the frame of reference that determines the normalization of the likelihoods, $\skew{-4}\tilde{\bL}\mapsto\bL$. Typically, as information is gained from data, the normalizing force of the frame of reference reduces the force of the same data in subsequent updates. 

All of this simply shows that Bayesian updating describes the change in probability distributions between two sets. That change between sets follows the universal principles given by the abstract Price equation. 

Prior work noted the analogy between natural selection and Bayesian updating \autocite{shalizi09dynamics,harper10the-replicator,campbell16universal}. Here, I emphasized a more general perspective that includes natural selection and Bayesian updating as examples of the common invariances and geometry that unify many topics. 

\section{Invariance and probability}

In the earlier section \textit{Affine invariance}, I showed that the Price equation is invariant to affine transformations $\bz\mapsto a+b\bz$. This section suggests that the Price equation's intrinsic affine invariance explains universal aspects of probability distributions in a more general and fundamental manner than Jaynes' focus on entropy and information.

The general form of probability distributions in \Eq{expDistn} followed from the constraint $\log\bq'=\log k-\Gl\bz$. Affine transformation does not change the force imposed by that constraint, because 
\begin{equation*}
  \log k-\Gl\bz\mapsto \log k - a\Gl -b\Gl\bz=\log k_a -\Gl_b\bz,
\end{equation*}
in which $k_a=ke^{-a\Gl}$ and $\Gl_b=b\Gl$. Because the constants, $k_a$ and $\Gl_b$, adjust to satisfy underlying constraints, the shift and stretch constants $a$ and $b$ do not alter the constraints or the final form of the probability distribution. 

Thus, the probability distribution in \Eq{expDistn}, arising from analysis of extreme action applied to a Lagrangian, is affine invariant with respect to $\bz$. We can make a more fundamental argument, by deriving the form of the probability distribution solely as a consequence of the intrinsic affine invariance of the Price equation.

In particular, shift invariance by itself explains why the probability distribution in \Eq{expDistn} has an exponential form \autocite{frank16common}. If we assume that the functional form for the probability distribution, $q_i=f(z_i)$, is invariant to a constant shift, $a+z_i$, then, dropping the $i$ subscripts and using continuous notation, by the conservation of total probability
\begin{equation}\label{eq:consTotal}
  \int k_0 f(z)\,\dd z = \int k_a f(a+z)\,\dd z=1
\end{equation}
holds for any magnitude of the shift, $a$, in which the proportionality constant, $k_a$, changes with the magnitude of the shift, $a$, independently of the value of $z$, in order to satisfy the conservation of total probability. 

Because $k_a$ is independent of $z$, the condition for the conservation of total probability is 
\begin{equation}\label{eq:basicShift}
 k_a f(a+z)=k_0 f(z).
\end{equation}
The invariance holds for any shift, $a$, so it must hold for an infinitesimal shift, $a=\Ge$. We can write the Taylor series expansion for an infinitesimal shift as
\begin{equation*}
 f(\Ge+z)=f(z) + \Ge f'(z)=\Gk_\Ge f(z),
\end{equation*}
with $\Gk_\Ge=1-\Gl\Ge$, because $\Ge$ is small and independent of $z$, and $\Gk_0=1$. Thus,
\begin{equation*}
 f'(z)=-\Gl f(z)
\end{equation*}
is a differential equation with solution
\begin{equation}\label{eq:exp}
  q=f(z)=ke^{-\Gl z},
\end{equation}
in which $k$ is determined by the conservation of total probability, and $\Gl$ is determined by $\zbar$. When $z$ ranges over positive values, $z>0$, then $k=\Gl=1/\zbar$. Invariance to stretch transformation by $b$ follows from the adjustment, $\Gl_b$, given above.

Affine invariance of the probability distribution with respect to $z$ implies additional structure. In particular, we can write $z=e^{\Gb w}$, in which a shift $w(z)\mapsto\Ga+w(z)$ multiplies $z$ by a constant, which does not change the form of the probability distribution. Thus, in terms of the shift-invariant scale, $w(z)$, we obtain the canonical expression that describes nearly all commonly observed continuous probability distributions \autocite{frank16common,frank16invariant}
\begin{equation}\label{eq:canonical}
  q\,\dd\psi=ke^{-\Gl e^{\Gb w}}\dd\psi,
\end{equation}
when we add a few additional details about the measure, $\dd\psi_z$, and the commonly observed base scales, $w(z)$. Understanding the abstract form of common probability patterns clarifies the study of many problems \autocite{frank16the-invariances,frank16invariant,frank18measurement} (see Appendix A).

\section{Meaning}

One cannot explain mathematical form by appeal to extrinsic physical notions. The structure of mathematical results does not follow from energy or heat or natural selection. Instead, those extrinsic phenomena arise as consistent interpretations for the structure of the mathematics. 

The mathematical structure can only be analyzed, explained and understood by reference to mathematical properties. For example, we may invoke invariance, conserved values, and geometry to understand why certain mathematical forms arise in the abstract Price equation description for changes in frequency, and why those same forms recur in many different applications. We may not invoke entropy or information as a cause, only as a description.

My goal has been to reveal the common mathematical structure that unifies seemingly disparate results from different subjects. The common mathematical structure arises primarily through simple invariances and their expression in geometry.
 
\section*{Acknowledgments}

\noindent The Donald Bren Foundation supports my research. I completed this work while on sabbatical in the Theoretical Biology group of the Institute for Integrative Biology at ETH Zürich.

%\vfill\eject

\mybiblio	% uses main.bib by default, add other bibs as needed

%%%%%%%%%%%%%%%%%%%%%%%%%%%%%%%%%%%%%%%%%%
\appendix

\section{Appendix A: Value of synthesis by invariance}

I have been asked to comment on how this synthesis of concepts may enhance scientific progress. The primary modes of progress follow two lines. 

First, one can more easily understand the vast literature that makes connections between disciplines. For example, information is often discussed as if it were a primary concept that clarifies the meaning of biological or physical principles. By contrast, in this synthesis based on the fundamental invariances expressed by the abstract Price equation, various information and entropy forms arise directly. This synthesis provides value if one feels curiosity about the similarity of mathematical forms or wishes to understand the literature that discusses such similarities.

Second, new mathematical results and new insights into empirical phenomena may follow. I believe this to be true. However, the argument for novel results and insights is nearly impossible to make. For any particular result or insight, it is always possible to claim that the same could have been achieved without the broader framing. Ascribing the origins of insight to a general framework is almost always subjective.

The strongest argument I can make arises from two personal anecdotes. It is only in these cases that I understand the origin of insight in relation to the broad use of invariance as a unifying perspective.

\subsection{Probability, invariance, and maximum entropy}

The first anecdote shows how observations in biology motivated my search for a broader synthesis of concepts between disciplines. That synthesis, in terms of invariance, helped me to understand the observed biological patterns. It also led to a unified understanding of the commonly observed probability distributions in terms of the invariances that define scale, and an understanding of the relations between the equations of thermodynamics, natural selection in biology, and probability patterns.

In my work on cancer and other aspects of age-related disease \autocite{frank07dynamics,frank16invariant}, I noted that a wide variety of seemingly different dynamical models of disease progression tended to converge to a few similar forms of probability distributions for the age of disease onset. At first, I used Jaynes' maximum entropy approach \autocite{jaynes57information,jaynes57informationII,jaynes03probability} to try and understand the relations between apparently complex processes and the resulting simple patterns \autocite{frank09the-common}. That worked, in the sense that one could find constraints that led to maximum entropy distributions that matched the data. 

The problem with maximum entropy is that the constraints simply describe the patterns in the data, without giving one a sense of how patterns arise and what relates different patterns to each other. Instead, one ends up with a catalog of the commonly observed probability distributions and the matching constraints for each distribution. 

Those difficulties led me to study the forms of commonly observed probability distributions. I felt that if I could understand probability patterns more deeply, I would be in a better position to understand the biological problems that interested me. And, along the way, I would perhaps better understand more general aspects of probability patterns.

Over many years, I developed a unified understanding of probability patterns in terms of invariance and scale \autocite{frank14how-to-read,frank16common}. I used that improved understanding of probability to enhance my analyses of age-related diseases \autocite{frank16invariant} and the size distributions of trees in forests \autocite{frank16the-invariances}. 

That work on invariance and scale in probability left open the puzzle of how that perspective related to Jaynes' classic maximum entropy approach. Although my invariance approach to probability patterns could stand separately from maximum entropy, Jaynes' approach was widely used and formed a standard against which my new work would reasonably be compared. Also, I developed my ideas by initially starting with maximum entropy, and Jaynes himself strongly hinted that invariance might be the way forward from where he left the subject \autocite{jaynes03probability}.

How could I connect my pure invariance approach to Jaynes' work on maximum entropy, which was developed explicitly as an extension to classical thermodynamics and statistical mechanics?

My work on probability seemingly has little relation to the Price equation. However, in my other studies, I had been using the Price equation as a tool to understand natural selection in biology \autocite{frank86hierarchical,frank95george,frank12naturalb}. Over time, I began to see the broader connections between the Price equation and information theory \autocite{frank09natural,frank12naturalc,frank13natural}. 

Through those studies of natural selection and the Price equation, I gained understanding of the dynamics of information. I was then able to see the connections between some of the classic results of thermodynamic change in entropy and the equations of natural selection. 

With that broader understanding of entropy and information dynamics, I could then synthesize Jaynes' maximum entropy approach to probability with my approach based on invariance and scale \autocite{frank17universal}. Some fundamental aspects of physical mechanics also began to fit within the unified structure \autocite{frank15dalemberts}. All of that abstract work fed back into my analyses and understanding of age-related diseases, the sizes of trees, and the distribution of enzyme rates \autocite{frank16invariant,frank16the-invariances}.

For any of the particular insights into empirical problems or any of the particular mathematical results, it would have been possible to achieve the same without a broader perspective or an attempt to unify between disciplines. However, in fact, the broader perspective and unification of disciplines played a primary role.

\subsection{The universal law of generalization in psychology}

The second anecdote shows how the broad framework led to a new insight for a particular discipline. In this case, I happened to read an article in \textit{Science} about an intriguing pattern in psychology \autocite{sims18efficient}. 

The probability that an organism perceives two stimuli as similar typically decays exponentially with the separation between the stimuli. The exponential decay in perceptual similarity is often referred to as the universal law of generalization \autocite{shepard87toward,chater03the-generalized}. 

Both theory and empirical analysis depend on the definition of the perceptual scale. For example, how does one translate the perceived differences between two circles with different properties into a quantitative measurement scale?

There are many different suggestions in the literature for how to define a perceptual scale. Each of those suggestions develops very specific notions of measurement based, for example, on information theory, Kolmogorov complexity theory, or multidimensional scaling descriptions derived from observations \autocite{chater03the-generalized,shepard87toward,sims18efficient}.

I showed that the inevitable shift invariance of any reasonable perceptual scale determines the exponential form for the universal law of generalization in perception \autocite{frank18measurement}. All of the other details of information, complexity, and empirical scaling are superfluous with respect to understanding why the universal law of generalization has the exponential form.

Certainly, the insight that the inevitable shift invariance of scale is a sufficient explanation does not require a broad conceptual framework derived from the Price equation. However, I was able to see immediately that solution only because I had for years been working toward a unified understanding of information, scale, and invariance. Many others had worked on this central puzzle in psychology without seeing the underlying simplicity.

\section{Appendix B: Mathematical expressions from various disciplines}

See Tables \ref{tbl:equations1} and \ref{tbl:equations2} on following pages.

\begin{table*}[]
\caption{Mathematical forms that highlight similarities between different disciplines, part 1}
\label{tbl:equations1}
\centering
%% \tablesize{} %% You can specify the fontsize here, e.g.,  \tablesize{\footnotesize}. If commented out \small will be used.
\begin{tabular}{llc}
\toprule
\textbf{Mathematical form}	& \textbf{Comments}	& \textbf{Equation}\\
\midrule

\tbi{Price equation:}\\[4pt]

\trow{\GD\zbar=\bdq\cdot\bz+\bq'\cdot\bdz}{Most general form; separates frequency, $\bq$, from property value, $\bz$; partitions frequency and property value change\ts}{price}
\trow{\GD\abar=\bdq\cdot\ba+\bq'\cdot\bda=0}{Canonical form; emphasizes conservation of total frequency; recover general form by coordinate change $\ba\mapsto\bz$\ts }{azero}
\midrule

\tbi{Mathematical relations:}\\[4pt]

\trow{\bdq\cdot\bz=\norm{\bdq}\norm{\bz}\cos\Gom}{Geometric equivalence for dot product; $\ba\equiv\bF$ yields abstract expression of physical work (see below)\ts}{work}
\trow{\bdq\cdot\bz=\cov(w,z)}{Equivalent statistical form}{cov}
\trow{\bq'\cdot\bdz=\E\lr{w\GD z}}{Equivalent statistical form}{secondterm}
\trow{\bdq\cdot\ba=\norm{\bdq\;/\sqrt{\bq}}^2}{Geometric expression for total distance between sets in terms of frequency; discrete generalization of Fisher information, $\F$\ts}{diverge}
\midrule

\tbi{Physical mechanics:}\\[4pt]

\trow{\GD\abar=\lr{\bF+\bI}\cdot\bdq=0}{Abstraction of D'Alembert's principle for physical work in conservative systems; work from direct forces, $\bdq\cdot\bF=\bdq\cdot\ba$, balances work from inertial forces, $\bdq\cdot\bI=\bq'\cdot\GD\ba$; generalize by coordinate transformation $\ba\mapsto\bz$; cases in which $\GD\zbar\ne0$ describe nonconservative systems\ts}{dalembert2}
\trow{\bdq\cdot\bF=\norm{\bdq}\norm{\bF}\cos\Gom}{Abstract form of work as distance moved, $\norm{\bdq}$, multiplied by component of force along path, $\norm{\bF}\cos\Gom$; for given lengths of force and frequency change vectors, the frequency changes that minimize the angle between force and frequency change maximize the work\ts}{work}
\midrule

\tbi{Information theory:}\\[4pt]

\trow{\bdq\cdot\bmm=\J\lr{\bq',\bq}}{Jeffreys divergence, $\J=\KL{\bq'}{\bq}+\KL{\bq}{\bq'}$ for $\bz\equiv\bmm=\log\bq'/\bq$\ts}{jeffreys}
\trow{\bdq\cdot\bmm\rightarrow\obq\cdot\ba}{For small changes, $\bmm\rightarrow\ba$ for $\bdq\rightarrow\obq$}{info}
\trow{\obq\cdot\ba=\norm{\obq\;/\sqrt{\bq}}^2=\F}{Abstract nondimensional expression of Fisher information as distance of relative frequency changes\ts}{fisherDist}
\trow{\norm{\obq\;/\sqrt{\bq}}^2=4\norm{\obr}^2=\F}{Fisher information as simple Euclidean geometric distance of frequency change in unitary coordinates, $\br=\sqrt{\bq}$\ts}{unitary}
\trow{\obq\cdot\bF=\obq\cdot\dd\log\bq=\F}{For $\bF\equiv\ba$, work of direct forces in terms of d'Alembert}{dalembertLog}
\trow{\obq\cdot\bI=\obq\cdot\dd\log^2\bq=-\F}{Work of inertial forces, the change in frame of reference}{dalembertLog}
\midrule

\tbi{Bayesian inference:}\\[4pt]

\trow{\log\bL\equiv\bmm;\; \bL-\bOne\equiv\ba}{For relative likelihood, $\bL$}{logL}
\trow{\qt'=\qt L_\Gth}{Bayesian updating}{bayesUpdate}
\trow{\bdq\cdot\log\bL=\J\lr{\bq',\bq}}{Follows from $\log\bL\equiv\bmm$}{bayesJ}
\trow{\bdq\cdot\log\bL\rightarrow\obq\cdot\ba=\F}{Follows from $\bmm\rightarrow\ba$ for $\GD\bq\rightarrow\obq$}{info}
\trow{\GD\skew{-4}\bar{\bL}=\GD\bq\cdot\bL+\bq'\cdot\GD\bL=0}{Likelihood form of canonical Price equation, $\bL-\bOne\equiv\ba$}{canonicalL}

\bottomrule
\end{tabular}
\end{table*}

\begin{table*}[]
\caption{Mathematical forms that highlight similarities between different disciplines, part 2}
\label{tbl:equations2}
\centering
%% \tablesize{} %% You can specify the fontsize here, e.g.,  \tablesize{\footnotesize}. If commented out \small will be used.
\begin{tabular}{llc}
\toprule
\textbf{Mathematical form}	& \textbf{Comments}	& \textbf{Equation}\\
\midrule

\tbi{Natural selection:}\\[4pt]

\trow{\GDF\mskip1mu\abar=\bdq\cdot\ba=V_w}{Natural selection moves population a distance equal to the variance in fitness; equivalent to abstract form of physical work with $\ba\equiv\bF$\ts}{Vw}
\trow{\GDF\mskip1mu\abar=V_w=V_g+V_\Ge}{Partition variance (distance) into part associated with genetic predictors, $V_g$, and part associated with other environment effects, $V_\Ge$\ts}{ftns}
\trow{\GD_{\raisebox{-2pt}{$\scriptstyle \mathrm{NS}$}}\mskip1mu\abar=V_g}{Analog of fundamental theorem, the part of total transmissible change caused by natural selection\ts}{ftns}
\trow{\GD\bar{p}=\cov(w,p)}{Replicator equation with $p\equiv z$ as gene frequency within individuals and $\bar{p}$ as population gene frequency\ts}{replicator}
\trow{\GD\bar{p}=\cov(w,p)+\E\lr{w\GD p}}{Group selection with $p\equiv z$ as gene frequency within groups, first term as selection between groups, and second term as selection within groups\ts}{groupSel}
\midrule

\tbi{Extreme action:}\\[4pt]

\trow{\LL=\obq\cdot\Gfb + \mathrm{constraints}}{Lagrangian as work of direct forces, $\Gfb\equiv\bF$; maximizing the work (action), $\obq\cdot\Gfb$, chooses the frequency changes, $\obq$, in the direction of the forces subject to constraints\ts}{actionV}
\trow{\oq_i=\Gk q_i \lr{\Gf_i-\Gfbar}}{Dynamics for constrained total frequency and constrained total distance, $\F=C^2$, with $\Gk=C/\Gs_\Gf$ and $\Gs_\Gf$ as standard deviation of forces\ts}{actionDyn}
\midrule

\tbi{Thermodynamics:}\\[4pt]

\trow{\ba=\GD\bq/\bq\rightarrow\obq/\bq}{Equivalence for small changes}{adefShort}
\trow{\bmm=\log\bq'/\bq\rightarrow\obq/\bq}{Define force $\Gfb\equiv\bmm$, with $q_i'=q_ie^{m_i} \rightarrow q_im_i$ }{malthus}
\trow{\obq\cdot\Gfb=\obq\cdot\log\bq'-\obq\cdot\log\bq}{Term $-\obq\cdot\log\bq$ is production of entropy}{splitForce}
\trow{\LL=-\obq\cdot\log\bq+\textrm{constraints}}{Maximizing Lagrangian maximizes production of entropy}{Lentropy}
\trow{\obq\cdot\log\bq'=-\Gl(\obq\cdot\bz)=-\Gl B}{If $\GD\bz=\mathbf{0}$, then constraint $\GD\zbar=B$ implies $\obq\cdot\bz=B$, which constrains vector of new frequencies, $\bq'$\ts}{dzconstraint}
\trow{\log\bq'=\log k-\Gl\bz}{Force of constraint in previous line}{dzconstraint}
\trow{\oq_i=\Gk q_i\lr{\EE_i^*-\Gl z_i^*}}{Dynamics that maximize entropy production}{entropyDyn}
\midrule

\tbi{Statistical mechanics:}\\[4pt]

\trow{q_i=ke^{-\Gl z_i}}{Solution for probability distribution from force of constraint at equilibrium, $\bq'=\bq$, and constraint $\zbar=\bq\cdot\bz=1/\Gl$\ts}{expDistn}
\trow{q_i=ke^{-\lr{z_i-\Gm}^2/2\Gs^2}}{Gaussian distribution from constraint $\Gs^2=\bq\cdot\lr{\bz-\Gm}^2$\ts}{expDistn}
\trow{q_i=ke^{-\Gl \tr(z_i)}}{Jaynesian maximum entropy distribution from constraint $\bq\cdot\tr(\bz)=1/\Gl$\ts}{expDistn}
\midrule

\tbi{Probability distributions:}\\[4pt]

\trow{q=ke^{-\Gl e^{\Gb w}}}{Canonical form of continuous probability distributions; $w(z)$ is shift-invariant scaling of $z$ such that probability pattern is invariant to constant shift, $w\mapsto\Ga+w$\ts}{canonical}

\bottomrule
\end{tabular}
\end{table*}

%\ifmulticol\end{multicols}\fi
\end{document}